\documentclass[aps,prl,twocolumn,superscriptaddress,showpacs,amsmath,amssymb,longbibliography]{revtex4-1}
\usepackage[english]{babel}
\usepackage{amsmath}
\usepackage{bm}
\usepackage{graphicx,bbm}
\usepackage{times}
\usepackage{epsfig}
\usepackage{soul}
\usepackage[utf8]{inputenc}
\usepackage[colorlinks,linkcolor=blue,citecolor=blue,urlcolor=blue]{hyperref}
\usepackage{color}
\usepackage{latexsym}
\usepackage{ulem}
\definecolor{nred}{RGB}{224,0,0}
\definecolor{nblue}{RGB}{28,130,185}
\definecolor{dgreen}{RGB}{78,138,21}
\definecolor{norange}{RGB}{230,120,20}
\makeatletter
\newcommand{\rmnum}[1]{\romannumeral #1}
\newcommand{\Rmnum}[1]{\expandafter\@slowromancap\romannumeral #1@}

\makeatother

\begin{document}

\title{Uniaxial stress effect on the quasi-one-dimensional Kondo lattice CeCo$_2$Ga$_8$}

\author{Kangqiao Cheng}
\affiliation{Wuhan National High Magnetic Field Center and School of Physics, Huazhong University of Science and Technology, Wuhan 430074, China}
\author{Binjie Zhou}
\affiliation{Wuhan National High Magnetic Field Center and School of Physics, Huazhong University of Science and Technology, Wuhan 430074, China}
\author{Cuixiang Wang}
\affiliation{Beijing National Laboratory for Condensed Matter Physics, Institute of Physics, Chinese Academy of Sciences, Beijing 100190, China}
\affiliation{School of Physical Sciences, University of Chinese Academy of Sciences, Beijing 100190, China}
\author{Shuo Zou}
\affiliation{Wuhan National High Magnetic Field Center and School of Physics, Huazhong University of Science and Technology, Wuhan 430074, China}
\author{Yupeng Pan}
\affiliation{Wuhan National High Magnetic Field Center and School of Physics, Huazhong University of Science and Technology, Wuhan 430074, China}
\author{Xiaobo He}
\affiliation{Wuhan National High Magnetic Field Center and School of Physics, Huazhong University of Science and Technology, Wuhan 430074, China}
\author{Jian Zhang}
\affiliation{Wuhan National High Magnetic Field Center and School of Physics, Huazhong University of Science and Technology, Wuhan 430074, China}
\author{Fangjun Lu}
\affiliation{Wuhan National High Magnetic Field Center and School of Physics, Huazhong University of Science and Technology, Wuhan 430074, China}
\author{Le Wang}
\affiliation{Shenzhen Institute for Quantum Science and Engineering, and Department of Physics, Southern University of Science and Technology, Shenzhen 518055, China}
\author{Youguo Shi$^{\dag}$}
\affiliation{Beijing National Laboratory for Condensed Matter Physics, Institute of Physics, Chinese Academy of Sciences, Beijing 100190, China}
\affiliation{School of Physical Sciences, University of Chinese Academy of Sciences, Beijing 100190, China}
\email[]{ygshi@iphy.ac.cn}
\author{Yongkang Luo$^*$}
\affiliation{Wuhan National High Magnetic Field Center and School of Physics, Huazhong University of Science and Technology, Wuhan 430074, China}
\email[]{mpzslyk@gmail.com}


\date{\today}

\begin{abstract}

Quantum critical phenomena in the quasi-one-dimensional limit remains an open issue. We report the uniaxial stress effect on the quasi-one-dimensional Kondo lattice CeCo$_2$Ga$_8$ by electric transport and AC heat capacity measurements. CeCo$_2$Ga$_8$ is speculated to sit in close vicinity but on the quantum-disordered side of a quantum critical point. Upon compressing the $\bm{c}$ axis, parallel to the Ce-Ce chain, the onset of coherent Kondo effect is enhanced. In contrast, the electronic specific heat diverges more rapidly at low temperature when the intra-chain distance is elongated by compressions along $\bm{a}$ or $\bm{b}$ axes. These results suggest that a tensile intra-chain strain ($\varepsilon_c >0$) pushes CeCo$_2$Ga$_8$ closer to a quantum critical point, while a compressive intra-chain strain ($\varepsilon_c <0$) likely causes departure. Our work provides a rare paradigm of manipulation near a quantum critical point in a quasi-1D Kondo lattice by uniaxial stress, and paves the way for further investigations on the unique feature of quantum criticality in the quasi-1D limit.

\end{abstract}


\maketitle

\begin{figure}[t]
\centering
\vspace*{0pt}
\hspace*{-25pt}
\includegraphics[width=10cm]{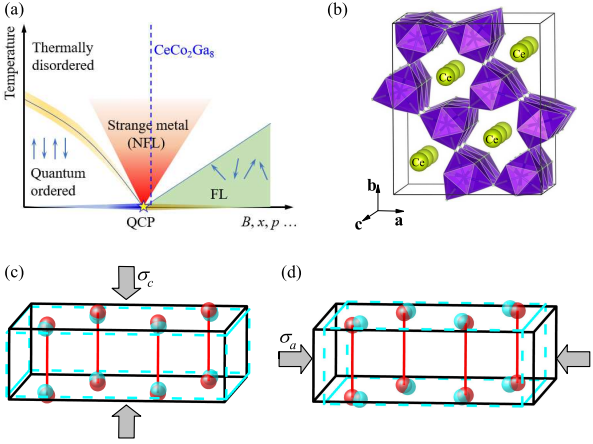}
\vspace*{-10pt}
\caption{\label{Fig1} (a) Schematic phase diagram. The blue dash line marks the position of CeCo$_2$Ga$_8$: on the right-hand side but in close vicinity of a QCP. (b) Crystalline structure of CeCo$_2$Ga$_8$. The cerium atoms form individual chains along $\bm{c}$ axis, surrounded by five CoGa$_9$ cages in the $\bm{ab}$ plane. (c-d), sketches of uniaxial stress effect. The red and cyan circles denote the atoms before and after stressing, respectively. When the stress is applied along $\bm{c}$, the intra-chain distance shortens at the expense of expansion in the $\bm{ab}$ plane; whereas for $\sigma_{a}$ (or $\sigma_{b}$), the intra-chain distance elongates. }
\end{figure}

Manipulation and control of the ground states near a quantum critical point (QCP) has attracted tremendous interests in recent decades \cite{Lohneysen-CeCu6Au2001,Grigera-Sr3Ru2O7QCP,Custers-YbRh2Si2QCP,Park-CeRhIn5QCP,LuoY-CeNiAsOQCP,JiaoL-CeRhIn5B,Ramshaw-YBCOSdH,Zhao-CePdAlQCP,Ran-UTe2FieldSC,Shen-CeRh6Ge4FMQCP,Worasaran-Ba122_CoNematicityQCP}. QCP, the point that separates the quantum-ordered and -disordered states on the phase diagram of a material, is generally achieved by terminating a phase transition \textit{continuously} at absolute zero, through the application of a certain non-thermal external control parameter, \textit{e.g.} magnetic field ($B$), chemical doping ($x$), physical pressure ($p$), etc [Fig.~\ref{Fig1}(a)]. In the vicinity of QCPs, many emergent quantum phenomena may appear, such as heavy-electron, strange metal [or non-Fermi liquid (NFL)], unconventional superconductivity, quantum spin liquid and so on. Heavy-fermion Kondo lattice compounds span a large material basis for exploring QCPs and investigating their unique nature \cite{Lohneysen-CeCu6Au2001,Custers-YbRh2Si2QCP,Park-CeRhIn5QCP,Custers-Ce3Pd20Si6QCP,LuoY-CeNiAsOQCP,JiaoL-CeRhIn5B,LuoY-CeNi2As2Pre,Zhao-CePdAlQCP,Shen-CeRh6Ge4FMQCP,Fuhrman-CeRu4Sn6QCP}. In this context, the ground state of the system is typically determined by a competition between  Kondo effect and Ruderman-Kittel-Kasuya-Yosida (RKKY) interaction\cite{Doniach}: while the RKKY exchange ($J_{RKKY}$) prefers a long-range magnetic ordering \cite{RKKY-RK,RKKY-K,RKKY-Y}, the Kondo effect ($T_K$) tends to screen and quench magnetic moments and thus stabilizes a non-magnetic ground state \cite{Hewson-Kondo}. A QCP is expected when $J_{RKKY}$ equals to $T_K$.

Natural questions concern whether a QCP can be realized in a one-dimensional (1D) or quasi-1D system, and what the nature of such QCPs is if they exist. They remain elusive, because (\rmnum{1}) it is generally believed that long-range magnetic order is hard to condensate in 1D systems \cite{Mermin-1D2DHeisenberg,Steiner-1DMagnetism}, the concept of QCP therefore seems ``meaningless"; (\rmnum{2}) Fermi liquid also breaks down in the 1D limit \cite{Luttinger_1963,Haldane_Luttinger}, so it is unclear what the ground state will be on the quantum disordered side; (\rmnum{3}) even in a relatively relax condition where long-range magnetic order might appear, \textit{i.e.} quasi-1D system, the Kondo coupling between local moments and conduction electrons is weakened due to the incomplete screening network, and furthermore, the inter-chain communication among single-ion Kondo singlets is also severely reduced \cite{Cheng-CeCo2Ga81D}, therefore, whether sufficiently strong Kondo effect can develop to compete against the RKKY interaction is still an open question; and (\rmnum{4}) so far most of the known heavy-fermion Kondo lattice are three-dimensional or quasi-two-dimensional, while examples of 1D or quasi-1D have been rare \cite{Krellner-YbNi4P2FMQCP,Shen-CeRh6Ge4FMQCP,Lyu-CeAu2In41D}. Extensive material basis and proper tuning on the existing candidates are both required to address these issues.

\begin{figure}[t]
\centering
\vspace*{-10pt}
\hspace*{-10pt}
\includegraphics[width=9cm]{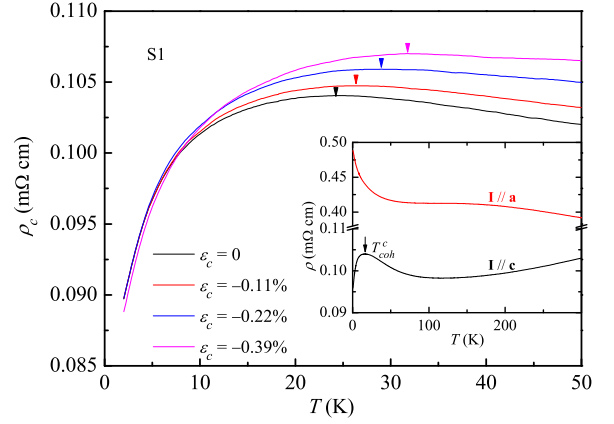}
\vspace*{-0pt}
\caption{\label{Fig2} Temperature dependent $\rho_c$ of CeCo$_2$Ga$_8$ measured under various $\varepsilon_{c}$. The arrows mark the onset of coherent Kondo temperature $T_{coh}^c$. The inset shows the profiles of $\rho_c(T)$ and $\rho_a(T)$ (data reproduced from Ref.~\cite{Cheng-CeCo2Ga81D}). Note that coherent Kondo scattering is seen only for $\bm{I}$$\parallel$$\bm{c}$.}
\end{figure}

Recently, Wang \textit{et al} reported the synthesis and physical properties of a new quasi-1D candidate Kondo lattice CeCo$_2$Ga$_8$ \cite{WangL-CeCo2Ga8}, which crystallizes in the YbCo$_2$Al$_8$-type orthorhombic structure\cite{Koterlin-Ce128}. Most interestingly, the cerium atoms in this compounds form individual chains along the $\bm{c}$ axis, and each chain is surrounded by five polyhedral CoGa$_9$ cages in the $\bm{ab}$ plane, cf Fig.~\ref{Fig1}(b). Since the inter-chain Ce-Ce distances (6.5 and 7.5 \AA) are much longer than the intra-chain distance (4.05 \AA), the compound is deemed as a candidate of quasi-1D Kondo lattice. Indeed, further resistivity measurements revealed that coherent Kondo scattering is only observed for electric current parallel to $\bm{c}$ ($\rho_c$), while both $\rho_a$ and $\rho_b$ remain incoherent down to 2 K, indicating the realization of \textit{Kondo chain} in CeCo$_2$Ga$_8$ \cite{Cheng-CeCo2Ga81D}, see also in the inset to Fig.~\ref{Fig2}. Specific heat, magnetic susceptibility and $\mu$SR measurements confirmed the absence of long-range magnetic ordering down to 70mK \cite{WangL-CeCo2Ga8,Bhattacharyya-CeCo2Ga8uSR}. Moreover, NFL behavior appears at low temperature as evidenced by the linear resistivity [$\rho_c(T) \sim T$] and logarithmic specific heat [$C/T\sim-ln T$] \cite{WangL-CeCo2Ga8}. At sub-Kelvin, $C/T$ tends to level off, suggesting that Fermi liquid potentially gets restored with a large renormalized effective mass at ultra-low temperature, and this is further supported by resistivity and specific heat measurements under pressure or magnetic field. Taken together, these features suggest that CeCo$_2$Ga$_8$ likely sits nearby but slightly on the quantum-disordered side of a QCP, seeing the blue dash line in the schematic phase diagram in Fig.~\ref{Fig1}(a). This provides a possible platform to investigate the nature of QCP in the quasi-1D limit.

In this work, we employed the uniaxial stress ($\bm{\sigma}$) as a control parameter to this quasi-1D Kondo lattice using a set of piezoelectric actuators \cite{Hicks-SRO214Strain}, and electric transport and thermodynamic properties were studied as a function of strain ($\bm{\varepsilon}$). Compared to conventional hydrostatic pressure effect that is essentially isotropic, the stress tuning possesses several advantages. First, because the stress effect is uniaxial, presumably it is more straightforward to change the intra-chain distance in such quasi-1D compounds and thus to tune the physical properties more efficiently. Second, one can apply either compressive or tensile stress by a controllable voltage, which enables \textit{in-situ} tuning either to approach or to depart the QCP.


Single crystalline CeCo$_2$Ga$_8$ was grown by a Ga self-flux method as described previously \cite{WangL-CeCo2Ga8}. The as-grown samples mostly are needle-like with typical length $\sim$3 mm along $\bm{c}$-axis, and about 1.5$\times$1.5 mm$^2$ in cross-section. Four samples (S1-S4) were prepared in this work. S1 and S2 were carefully polished to make the long side along $\bm{c}$- and $\bm{a}$-axes, for measurements of electrical resistivity $\rho_c$ and $\rho_a$, respectively. The forces were applied uniaxially along the electric current by Razorbill Instruments Cryogenic stress cell (FC100), and the magnitude of stress is measured using a pre-calibrated capacitive dilatometer. Heat capacity of CeCo$_2$Ga$_8$ under stress was measure by an AC calorimetric method \cite{Wilhelm-ACCalorimetry,Li-CacStress} on samples S3 and S4, and the stress was applied along $\bm{a}$ and $\bm{b}$ axes, respectively. Chromel-Au$_{99.93\%}$Fe$_{0.07\%}$ thermocouple was used to measure the heat-temperature response \cite{Kohler-Thermocouple,Stockert-CrAuFe}.


We start from resistivity measurements with compressive stress applied in $\bm{c}$ axis, which is the same direction as Ce-Ce chain. In this configuration, a compressive stress shortens the intra chain Ce-Ce distance, but slightly increases the inter chain distances. This can be seen from the Hooke's law for a crystal:
\begin{equation}
\bm{\sigma}=\bm{C}\cdot\bm{\varepsilon},
\label{Eq1}
\end{equation}
where $\bm{\sigma}$ is the stress tensor, $\bm{C}$ is the elastic moduli tensor, and $\bm{\varepsilon}$ is the strain tensor. We obtained the elastic moduli by Resonant Ultrasound Spectroscopy (RUS) measurements \cite{Migliori-RUS},
\begin{equation}
\begin{aligned}
\bm{C}(\text{GPa})=\left[
  \begin{array}{ccc}
        172.46~ & ~ 55.34~ &  ~ 53.11 \\
         55.34~ & ~161.55~ &  ~ 69.56 \\
         53.11~ & ~ 69.56~ &  ~169.96
\end{array}
\right],
\end{aligned}
\label{Eq2}
\end{equation}
where the irrelevant shear moduli are not shown here. More details about the RUS results on CeCo$_2$Ga$_8$ will be published separately \cite{ZhouB-CeCo2Ga8RUS}. With these parameters, we are able to convert the stress into strain as labeled in the legends of Fig.~\ref{Fig2}. Note that the sign ``$-$" in strain means compression.

As the intra-chain distance shortens, naively, one expects that the communications among single-ion Kondo singlets strengthens and thus the coherent Kondo effect enhances. However, on the other hand, since the RKKY interaction $J_{RKKY}\propto cos(2k_F r)/r^3$ (where $r$ is the distance between local moments, and $k_F$ is Fermi wave vector) \cite{RKKY-RK,RKKY-K,RKKY-Y}, it is hard to predict who will increase faster with strain. If Kondo effect wins, the system should move further into the quantum-disordered region; otherwise, it should approach closer to QCP. Figure \ref{Fig2} displays the temperature dependence of $\rho_c$ under various strains. At ambient, $\rho_c(T)$ initially decreases upon cooling, and then turns up below $\sim$100 K where the incoherent Kondo effect sets in (see the inset to Fig.~\ref{Fig2}). A broad peak forms at about 24 K, characteristic of the onset of coherent Kondo scattering and the development of renormalized heavy electron bands \cite{WangL-CeCo2Ga8}. The coherent Kondo temperature is defined as the position where $\rho_c$ maximizes. It should be mentioned that the $T_{coh}^c$ in this work is relatively higher than previously reported \cite{WangL-CeCo2Ga8,Cheng-CeCo2Ga81D}, probably due to sample quality dependence. One then clearly finds that $T_{coh}^c$ increases monotonically with compressive strain $\varepsilon_c$, with an increasing rate $\sim$ 20 K/\%. This manifests that the Kondo effect dominates in this process, and thus CeCo$_2$Ga$_8$ moves farther away from QCP. In other words, the correct way to tune it to QCP should be to stretch it along $\bm{c}$. However, we noticed that this crystal is rather brittle to tension, even a little tensile force causes the $\rho_c$ measurements to fail.

\begin{figure}[t]
\centering
\vspace*{-10pt}
\hspace*{0pt}
\includegraphics[width=9cm]{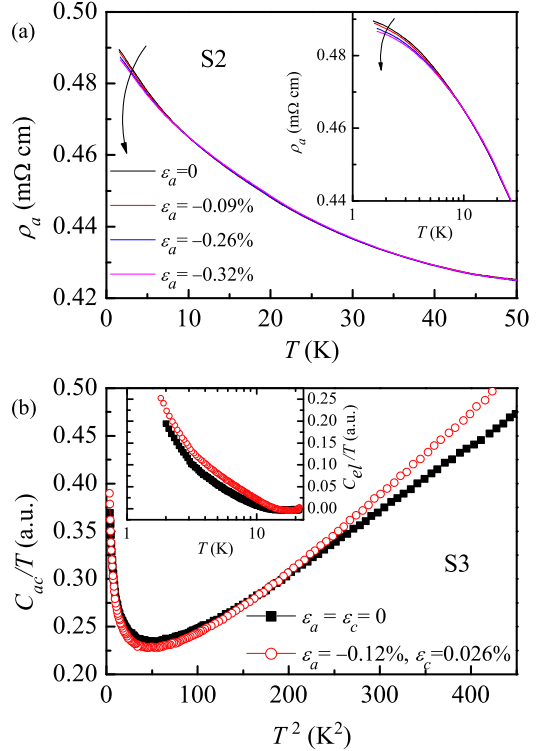}
\vspace*{-10pt}
\caption{\label{Fig3} (a) Temperature dependence of $\rho_a$ measured under various $\varepsilon_{a}$. (b) AC heat capacity divided by $T$, plotted vs. $T^2$, measured at $\varepsilon_{a}$=0 and $-$0.12\%. The inset shows $C_{el}/T$ plotted in the $\log T$ scale. }
\end{figure}

An alternate attempt is to compress the crystal within the $\bm{ab}$ plane, and presumably this is to elongate the intra-chain distance. In this configuration, the intra-chain coupling is expected to be weakened at the expense of some enhancement in inter-chain coupling. We assume the latter is less crucial. This idea is firstly testified by the $\rho_a$ measurements with stress applied in $\bm{a}$ axis. (For technical reason, it is difficult to measure $\rho_c$ in this configuration.) The ratio of $\varepsilon_c$ and $\varepsilon_a$ is determined by the Poison's ratio $\nu_{13}$$\equiv$$-S_{13}/S_{11}$$\approx$0.21, where $S_{11}$ and $S_{13}$ are the elements of elastic compliance tensor $\bm{S}$,
\begin{equation}
\begin{aligned}
\bm{S}(\text{GPa}^{-1})=\bm{C}^{-1}=\left[
  \begin{array}{ccc}
        6.79~ & ~-1.71~ &  ~ -1.42 \\
       -1.71~ & ~ 7.95~ &  ~ -2.72 \\
       -1.42~ & ~-2.72~ &  ~7.44
\end{array}
\right]\times 10^{-3}.
\end{aligned}
\label{Eq2}
\end{equation}
Figure \ref{Fig3}(a) displays $\rho_a$ as a function of $T$ measured at different $\varepsilon_{a}$. Unlike $\rho_{c}(T)$, the resistivity for $\bm{I}$$\parallel$$\bm{a}$ is semiconducting-like without any trace of Kondo coherence (see the inset to Fig.~\ref{Fig2}). We noticed that all the $\rho_a(T)$ curves essentially overlap only except for below $\sim$6 K where $\sigma_a$ weakly suppresses $\rho_a$. This implies that compressing along $\bm{a}$ tends to promote coherent Kondo scatting in $\bm{a}$, which is not surprising. However, we should also point out that the uniaxial stress effect along $\bm{a}$ is much weaker than along $\bm{c}$, manifesting that elastic-electronic coupling and thus the tunability are very anisotropic, and this provides additional evidence for the quasi-1D nature of CeCo$_2$Ga$_8$.

\begin{figure}[t]
\centering
\vspace*{-10pt}
\hspace*{0pt}
\includegraphics[width=9cm]{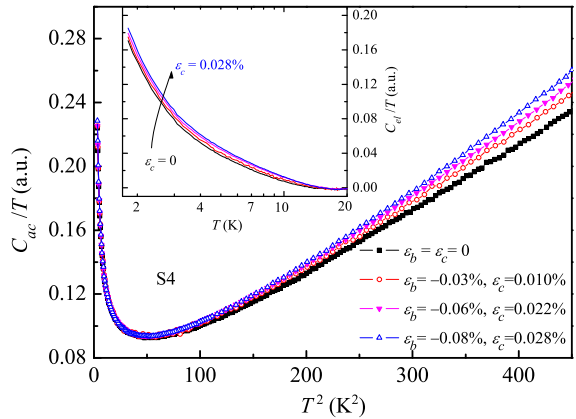}
\vspace*{-10pt}
\caption{\label{Fig4} AC heat capacity divided by $T$, plotted vs. $T^2$, measured in the presence of $\sigma_b$. The inset shows $C_{el}/T$ plotted in the $\log T$ scale. }
\end{figure}

Since resistivity is a tensor quantity, $\rho_a$ does not reflect much information about the intra-chain electronic correlation effect, we therefore turn to the bulk property measurements, \textit{viz.} AC heat capacity ($C_{ac}$), and the results are shown in Fig.~\ref{Fig3}(b). This provides a \textit{semi-quantitative} measure for the strain dependent electronic correlation. Here we compare the data for $\varepsilon_a$=0 and $-$0.12\%, and the latter corresponds to a small expansion strain along $\bm{c}$, $\varepsilon_c$=0.026\%. Between 10 and 20 K, $C_{ac}/T$  is linear in $T^2$ due to the phonon contribution \cite{Ashcroft-SSP}. A notable feature is that the slope of $C_{ac}/T$ vs. $T^2$ increases and hence Debye temperature decreases in the presence of $\sigma_a$. We obtain the phonon contribution by fitting $C_{ac}$ to a Debye $T^3$ law between 10 and 20 K, and extrapolate this to the lower $T$ range. After subtracting this phonon contribution, the electronic contribution to specific heat, $C_{el}/T$, is displayed in the inset to Fig.~\ref{Fig3}(b). Apart from a plausible upturn below 3 K which is due to a calibration issue of the Chromel-Au$_{99.93\%}$Fe$_{0.07\%}$ thermocouple used \cite{Kohler-Thermocouple,Stockert-CrAuFe}, the most prominent feature is that $C_{el}/T$ diverges more rapidly at low temperature under $\varepsilon_c$=0.026\%. This signifies that the effective mass of quasiparticles is further increased upon tensile $\varepsilon_c$, and is in agreement with the behavior when approaching a QCP (\textit{e.g.} see review \cite{Lohneysen-RMP2007}). It should be noted that at present we can not fully exclude the possibility that the rapid increase in $C_{el}/T$ at low temperature originates from an underlying magnetic ordering whose transition temperature is below our base temperature $\sim$1.7 K. Such a situation might appear if the system has been over-tuned to the quantum-ordered phase [cf Fig.~\ref{Fig1}(a)]. However, this scenario is not very likely considering the rather small $\varepsilon_c$ we have reached in this experiment, and no magnetic transition was observed down to 70 mK at ambient\cite{WangL-CeCo2Ga8,Bhattacharyya-CeCo2Ga8uSR}. Sub-Kelvin measurements are needed to clarify this problem. Similar trend is also observed when the uniaxial stress is applied along $\bm{b}$ axis, see Fig.~\ref{Fig4}.

According to the Global Phase Diagram theory \cite{SiQ-PhysB2006,Custers-Ce3Pd20Si6QCP}, the quantum critical points in heavy-fermion materials are generally classified in two types: a conventional spin-density-wave (SDW) type QCP \cite{Hertz-QCP,Millis-QCP}  and an unconventional Kondo-destruction type QCP \cite{SiQ-localQCP,Coleman-QCP2005,Gegenwart2008}. Across the Kondo-destruction type QCP, accompanied with the disappearance of long-range magnetic ordering, the 4$f$ electrons undergo a localized-delocalized transition, therefore, the Fermi surface topology changes sharply \cite{Shishido-CeRhIn5dHvA,Paschen-YbRh2Si2Hall,LuoY-CeNiAsOQCP,JiaoL-CeRhIn5B}. Other important features of the Kondo-destruction type QCP include an $\omega/T$ scaling of dynamic spin susceptibility, a modified Curie-Weiss form of static spin susceptibility, and strange metal (NFL) behavior with divergent quasiparticle effective mass \cite{Schroder-CeCu6AuFluc,Stockert-CeCu62DFluc,Schroder-CeCu6AuQCP,SiQ-localQCP}. Since Kondo destruction generically requires large a frustration parameter and spin fluctuations, it thus favors lower dimensionality \cite{Custers-Ce3Pd20Si6QCP}. Whether these phenomena will also appear in this quasi-1D system needs further confirmation. In particular, different kind of magnetic fluctuations (ferromagnetic (FM) or antiferromagnetic (AFM)) with different dimensionalities (2D or 3D) lead to different sorts of NFL behaviors, characterized by the different forms of temperature dependent measurables like resistivity, spin susceptibility, specific heat, spin-lattice relaxation rate and so on. For instance, in Moriya and Takimoto's theory \cite{Stewart-RMPNFL}, 2D and 3D AFM fluctuations yield specific heat $C_{m}/T$ obeying $-\log T$ and $\gamma_0-aT^{1/2}$, respectively, while 2D FM fluctuations result in $C_{m}/T$$\sim$$T^{-1/3}$ \cite{Moriya-QCP}. Slightly away from QCP, in the Fermi-liquid regime, $C_{m}/T$ conforms to $\log \frac{1}{r}$, $-r^{1/2}$ and $r^{1/2}$ for 2D-AFM, 3D-AFM and 2D-FM types of QCPs, respectively, where $r$$\equiv$$\delta$$-$$\delta_c$ parameterizes the ``distance" to QCP (refer to \cite{Zhu-Gruneisen} for more details). How the strange metal behaves near a QCP in the quasi-1D limit remains unclear. The stress-tuned CeCo$_2$Ga$_8$ thus provides a new access to these peculiar quantum critical phenomena. In our AC heat capacity experiment (which is semi-quantitative), we confirmed that $C_{el}/T$ increases nearly logarithmically between 3 and 10 K, and this trend retains when approaching QCP (see the insets to Figs.~\ref{Fig2} and \ref{Fig3}). Because of some uncertainty in the measurements below $\sim$3 K as mentioned here above, it is premature to draw a full conclusion before more precise measurements down to lower temperatures can be done. Furthermore, it is also unknown whether ferromagnetic or antiferromagnetic long-range order will be established if the system is over-tuned across the QCP. More systematic physical-property measurements at milli-Kelvin range in the presence of even more powerful stress cell are required in the future. Some relevant works have been on the way.


On the example of quasi-1D heavy-fermion Kondo lattice CeCo$_2$Ga$_8$, we systematically studied the uniaxial stress effect by anisotropic transport and thermodynamic properties measurements. The results manifest that a tensile intra-chain strain ($\varepsilon_c >0$) pushes CeCo$_2$Ga$_8$ closer to a quantum critical point, while a compression intra-chain strain ($\varepsilon_c <0$) likely causes departure. Our work provides a rare paradigm of manipulation near a quantum critical point in a quasi-1D Kondo lattice by uniaxial stress, and paves the way for further investigations on the unique feature of quantum criticality in the quasi-1D limit.

\section{Acknowledgments}

Y. Shi acknowledges Beijing Natural Science Foundation (Z180008) and K. C. Wong Education Foundation (GJTD-2018-01).


%

\end{document}